\documentclass[a4paper,fleqn,usenatbib]{mnras}

\usepackage[T1]{fontenc}
\usepackage{ae,aecompl}
\usepackage{lipsum}
\usepackage{mathtools}
\usepackage{cuted}
\usepackage{float}

\usepackage[authoryear]{natbib}

\newcommand{\demsort}[2]{#2}

\usepackage{graphicx}	
\usepackage{amsmath}	
\usepackage{amssymb}	
\usepackage{breqn}

\usepackage{multirow}
\usepackage{booktabs}
\usepackage{amsmath}
\usepackage{hyperref}
\usepackage{amssymb}

\title[On the formation of globular clusters]{On the formation of globular clusters: comparison with observations}

\author[Jim\'enez et al.]{
Santiago Jim\'enez,\thanks{E-mail: sjimenez@inaoep.mx}
Guillermo Tenorio-Tagle,
Sergiy Silich
\\
Instituto Nacional de Astrof\'isica, \'Optica y Electr\'onica, AP 51, 72000 Puebla, M\'exico\\
}

\date{Accepted XXX. Received YYY; in original form ZZZ}

\pubyear{2022}

\defcitealias{2021MNRAS.505.4669J}{paper I}

\begin{document}
\label{firstpage}
\pagerange{\pageref{firstpage}--\pageref{lastpage}}
\maketitle

\begin{abstract}
The paper deals with the conditions required to form at least two stellar generations in globular clusters under the constraints generated by feedback from massive stars as well as radiative cooling and the metallicity of the primordial clouds. Our calculations are based on two main constraints to the star formation efficiency of the first stellar generation (1G) $\epsilon_{1G}$. First, $\epsilon_{1G}$ is restricted to warrant that stellar winds and supernovae do not disrupt the leftover gas out of which a second generation (2G) would form. Second, $\epsilon_{1G}$ is also limited such that the metallicity enhancement caused by trapped supernovae is, in agreement with the observations, not larger than  $\sim$ 0.1 dex. Several central parameters define the globular clusters end result: the mass and radius of the primordial clouds, their metallicity and $\epsilon_{1G}$. The parameter space composed by models which fulfilled all constraints, is here shown to coincide remarkably well with the scattered observed anti-correlation between the fraction of first generation stars ($f_{\textrm{1G}}$) and total cluster mass. Our models also discern, in agreement with the data, between single and multiple population clusters in a metallicity versus mass (or radius) plane. Hence, our results suggest that the presence of multiple stellar populations is closely linked to the ability of proto-globular clusters to retain a fraction of leftover gas. 

\end{abstract}




\section{Introduction}\label{introductionI}
Most Galactic globular clusters (GGCs) are not simple stellar populations (SSPs) but instead are complex objects, hosting at least two different stellar populations characterized mainly by differences in their light element compositions \citep[e.g.][]{ 1999Natur.402...55L, 2004ApJ...605L.125B, 2009A&A...505..139C, 2012A&A...544A..12G, 2015A&A...578A.116C, 2015AJ....149...91P,2015MNRAS.447..927M, 2015ApJ...808...51M, 2019MNRAS.487.3815M}. The multiple populations (MPs) phenomenon was also found in extra-galactic systems, such as in the old \citep[e.g.][]{2009ApJ...695L.134M, 2016ApJ...829...77D,2019MNRAS.486.5581G} and intermediate age ($2-6$ Gyr old) Magellan Clouds (MCs) globular clusters (GCs) \citep[e.g.][]{2007AJ....134.1813M, 2017MNRAS.465.4159N}, in GCs located in other local dwarf galaxies \citep[e.g.][]{2014ApJ...797...15L, 2019MNRAS.490L..67S, 2021A&A...648A..70F} and in the GCs of M31 \citep{2013ApJ...776L...7S}. 

While multiple stellar populations are ubiquitous among globular clusters, no clusters are alike regarding their detailed chemical compositions \citep{2015MNRAS.454.4197R}. Hence, one way to elucidate clues about the origin of MPs is to study the dependence of this phenomenon with global parameters such as the metallicity, mass, compactness and escape velocity \citep[e.g.][]{2010A&A...516A..55C, 2017MNRAS.464.3636M,2018MNRAS.478.1520B, 2019AJ....158..202L, 2019A&A...624A..24C, 2020MNRAS.491..515M, 2021MNRAS.505.2548M}. 

The Hubble Space telescope (HST) UV legacy survey \citep{2015AJ....149...91P} provided high-accuracy photometry of a large sample of GCs. To analyze the data, \cite{2012ApJ...744...58M} introduced a pseudo color-color diagram (known as chromosome map), able to recognize  different stellar populations
within GCs since the pseudo colors are sensitive to differences in light elements compositions. With the chromosome maps, the number of first generation (1G) stars, which are those with similar light elements abundances as halo stars of the same metallicity, can then be calculated. The fraction of 1G stars ($f_{\textrm{1G}}$) with respect to the total number of stars was found to be anti-correlated with the total cluster mass \citep{2017MNRAS.464.3636M}: typically $f_{\textrm{1G}} \sim 0.2$ for massive clusters ($M>10^{6}$ M$_{\odot}$) while $f_{\textrm{1G}} \gtrsim 0.6$ for low mass clusters ($M < 3 \times 10^{5}$ M$_{\odot}$). This result, put together with the fact that lower mass open clusters (OCs) do not show evidence for MPs \citep[e.g.][]{2012A&A...548A.122B,2015MNRAS.446.3556M}, hints to the existence of a critical mass triggering the MPs within globular clusters. Such a transition mass, if it exists, would put an additional strong constraint which may enlighten the origin of the MPs within globular clusters.

From a theoretical perspective, the anti-correlation of  $f_{\textrm{1G}}$ with cluster mass and whether or not there is a triggering mass for the onset of MPs within GCs, should be explained by a consistent GCs formation mechanism. The most popular models are self-enrichment scenarios, that put forward different type of stars in different evolutionary stages, as the sources of enriched material out of which secondary populations (2G) formed. Examples of possible polluters are AGB stars \citep[e.g.][]{1981ApJ...245L..79C, 2004ApJ...611..871D, 2010MNRAS.407..854D, 2016MNRAS.458.2122D}, interacting binaries \citep[e.g.][]{2009A&A...507L...1D, 2013MNRAS.436.2398B, 2019ApJ...879...58T}, massive stars \citep[e.g.][]{2007A&A...464.1029D, 2007A&A...475..859D, krause2013superbubble, 2017ApJ...835...60W, 2019ApJ...871...20S}, very massive stars  \citep[VMSs, e.g.][]{2014MNRAS.437L..21D, 2018MNRAS.478.2461G}, and mixes of polluters such as massive and AGB stars \citep[e.g.][]{2018ApJ...869...35K} and a VMS and AGB stars \citep[e.g.][]{2019MNRAS.485.4311J}. These formation scenarios provide H-burning products that are required to explain the chemical signatures of MPs within GCs \citep[e.g.][]{1999A&A...347..572A, 2016EAS....80..177C}, but fail to explain simultaneously the increasing amount of data  (see recent reviews by \cite{2018ARA&A..56...83B} and \cite{2019A&ARv..27....8G}). 

There are also scenarios that contemplate total disruption of the left over gas due to stellar feedback, with the need of a later accretion of more primordial gas to end up with another stellar generation (as in \citealt{2008MNRAS.391..825D} , \citealt{2011MNRAS.415.1304D}, \citealt{2015ApJ...814L..14C}, \citealt{2016MNRAS.461.4088D}, \citealt{2019MNRAS.489.3269C}) or invoking a rapid cooling of the matter reinserted by stellar winds and supernovae. This driven by the local gravity may collapse and lead to a second stellar generation. Indeed, if radiative cooling is taken into account, gas expulsion by a global wind may be considerably reduced \citep[e.g.][]{2003ApJ...590..791S, 2007ApJ...658.1196T, 2007A&A...471..579W,2014ApJ...792..105P,2014MNRAS.442.2701R,2016MNRAS.456L..20B, 2017ApJ...835...60W,2017MNRAS.470..977L,2018MNRAS.478.2794N,2020MNRAS.499..748D}. 

Lithium (Li) is efficiently destroyed at temperatures larger than $\sim 2.5 \times 10^{6}$ K, which then implies that H-burning products should be Li-free given the higher operation temperatures of p-capture reactions \citep{2002PASP..114..375S}. Therefore, the observed presence of Li within 2G stars \citep[e.g.][]{2010ApJ...716L.166D, 2011MNRAS.412...81M, 2012A&A...539A.157M,2014ApJ...791...39D, 2015MNRAS.449.4038D} suggest that these stars cannot be formed only from the 1G stellar ejecta. An amount of dilution with pristine gas (i.e., the gas from which the 1G was formed) is necessary to account for the Li abundances \citep{2006A&A...458..135P}. Although AGB stars could produce Li (unlike massive stars and VMSs) through the Cameron-Fowler mechanism \citep[e.g.][]{1971ApJ...164..111C, 2002A&A...393..215V}, the AGB scenario still requires dilution in order to convert the Na-O correlation (as given by yields from AGB evolution models) into the observed Na-O anticorrelation \citep[e.g.][]{2011MNRAS.415.1304D}.

Furthermore, most GCs are mono-metallic regarding heavy elements, with iron spreads between cluster stars typically within $\Delta \textrm{[Fe/H]} \leq 0.1$ dex \citep{2009A&A...508..695C,2019ApJS..245....5B}. Hence, the ejecta from Type II supernovae, enriched in iron-peak elements, is usually discarded as part of the polluter material (the supernova (SN) avoidance problem, e.g. \citealt{2015MNRAS.454.4197R}, \citealt{2020IAUS..351..251M}). 

Although it has been generally assumed that none of the SN ejecta is retained and used as part of the polluter material, in some models GCs can retain a fraction of this ejecta such as in the self-enrichment models by \cite{2018ApJ...863...99B, 2021arXiv211014571B}. \cite{2021MNRAS.506.4131W} uses the observed iron (Fe) spreads to put constraints on the time-scales of the star formation episodes of young clusters and also found that for a given IMF, between $\sim 0.5 \%$ and $\sim 70 \%$ of the expected SN must be retained in order to explain the iron spreads of the clusters in their sample. \cite{2021MNRAS.505.4669J} (hereafter referred to as \citetalias{2021MNRAS.505.4669J}) presents calculations that include stellar winds and supernovae feedback.The initial conditions of the proto-clusters and the 1G star formation efficiency are constrained in order to obtain iron spreads below the observational limit and also to allow for a fraction of the leftover gas to survive the stellar feedback. Hence, in \citetalias{2021MNRAS.505.4669J} models, we have assumed that the first burst of star formation is self-limited as the leftover gas is impacted by the stellar feedback (winds and supernovae) even if a global star cluster wind is suppressed. In such cases stellar feedback enhances the cloud velocity dispersion and inhibits a further collapse until massive stars disappear from the scene (about  $\sim 40$ Myr later). In this way, 2G stars could then form from retained pristine gas with small iron spreads.

Here, following \citetalias{2021MNRAS.505.4669J}, we conduct a detailed study of the impact caused on the final cluster appearance due to the host cloud mass, metallicity, size and 1G star formation efficiency. Our calculations are compared with recent observational results which display the fraction of first generation stars $f_{\textrm{1G}}$  as a function of the total star cluster mass and accommodate the observed clusters in the metallicity and age versus cluster mass or radius parameter space. Such diagrams indicate the conditions required to form MPs clusters.

The paper is organized as follows: Section 2 presents a full description of our model, including the assumed physics of stellar feedback while Section 3 remarks on the constraints imposed by both stellar winds and supernova explosions. It is shown that the ability of the host molecular cloud to form clusters with MPs is determined by the set of input parameters: the 1G mass, radius, metallicity and the 1G star formation efficiency ($\epsilon_{\textrm{1G}}$). Section 4 shows that our results are in agreement with data in the 1G fraction ($f_{\textrm{1G}}$) vs total initial stellar mass ($M_{\textrm{ini}}$) plane. Section 5 discusses the impact of the metallicity on the formation of MPs. Our models are shown to split MPs from SSPs in  the $\textrm{[Fe/H]}$ vs mass (and radius) plane. Finally, Section 6 presents a summary and conclusions from our results.

\section{Model Setup}
\subsection{Initial mass distribution}

It is assumed that the first stellar generation (1G) within a proto-cluster cloud of mass $M_{\textrm{tot}}$ forms with a mass $M_{\textrm{1G}}=\epsilon_{\textrm{1G}} M_{\textrm{tot}}$, where $\epsilon_{\textrm{1G}}<1$ is the star formation efficiency. Hence, the 1G stars initially are embedded into the leftover gas with mass $M_{\textrm{gas}}=\left(1-\epsilon_{\textrm{1G}} \right) M_{\textrm{tot}}$. 

Following \citetalias{2021MNRAS.505.4669J} we assume a Gaussian density distribution with core radius $R_c$ for the stars and leftover gas and that the 1G fully samples a Kroupa initial mass function (IMF). No further consideration was made about  mass segregation as its impact on our main results is negligible (see Appendix \ref{Ap1}). It is also assumed that the 1G stars and pristine gas have the same metallicity. The cloud pressure is determined by the equation of hydrostatic equilibrium \citep[e.g.][]{2015ApJ...814L..14C}:

\begin{equation}\label{eq1}
\frac{d P_{\textrm{g}}}{dr}=-\frac{G M \left(r \right) \rho_{\textrm{g}} }{r^{2}},
\end{equation}
where:
\begin{equation}\label{eq2}
\rho_{g} \left(r \right)=\left\{
	\begin{array}{ll}
		\frac{M_{\textrm{gas}}}{\left( 2 \pi\right)^{3/2}R_{\textrm{c}}^{3}} \exp \left[-\frac{1}{2}\left( \frac{r}{R_{\textrm{c}}}\right)^{2} \right],  & \mbox{if }  r \leq  R_{\textrm{SC}},\\
     \rho_{\textrm{amb}} & \mbox{if } r > R_{\textrm{SC}},
	\end{array}
\right.
\end{equation}
is the gas density, G is the gravitational constant, $M\left(r\right)$ is the total mass contained within radius $r$  and $\rho_{\textrm{amb}}$ is the uniform density of the surrounding medium.

\subsection{Stellar Winds}\label{winds}

Winds driven by massive stars sweep up their surrounding gas into thin shells. If neighboring bubbles collide with each other, the shock-heated gas merge leading to a large overpressure in the central zone of the cluster. In such cases a global cluster wind that expels the leftover gas from the cluster is formed  \citep[e.g.][]{1985Natur.317...44C, 2000ApJ...536..896C}. 
\begin{figure}
\centering
	\includegraphics[width=1.0\columnwidth]{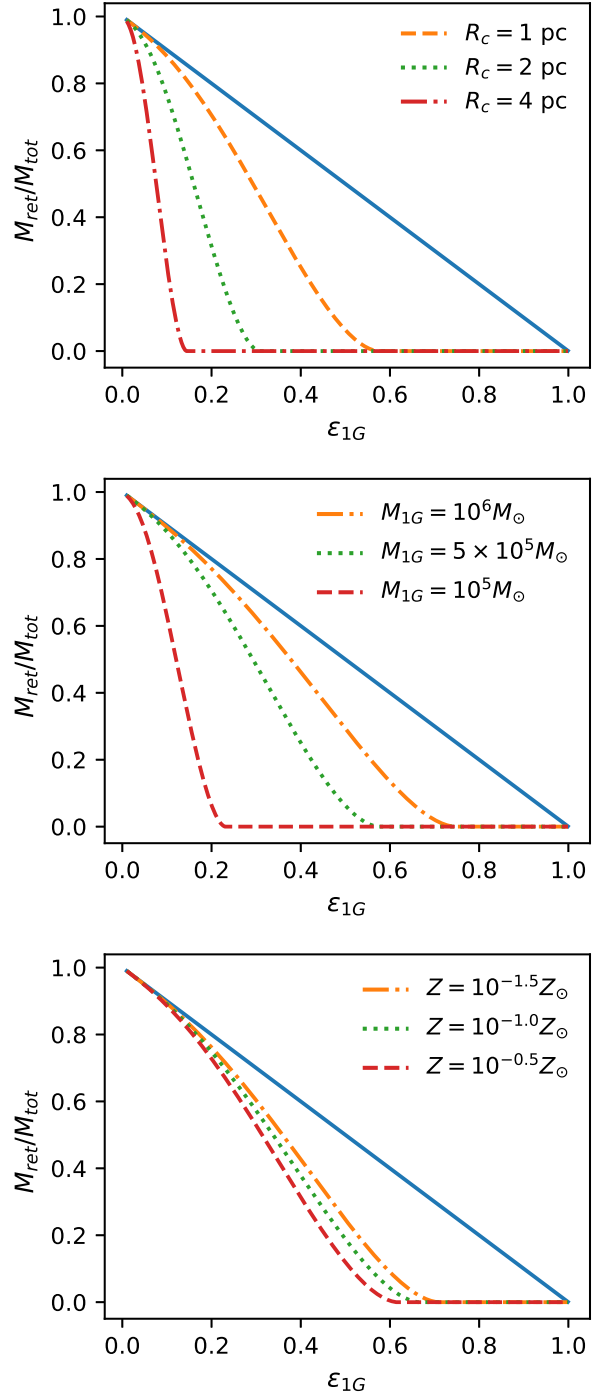}
        \caption{Upper panel: retained gas to the initial cloud mass ratio $M_{\textrm{ret}}/M_{\textrm{tot}}$ versus the star formation efficiency $\epsilon_{\textrm{1G}}$ for different core radii (see the legend) in the case when $Z=Z_{\odot}$ and $M_{\textrm{1G}}=5 \times 10^{5}$ M$_{\odot}$. Middle panel: the same but for different values of the 1G mass and fixed core radius $R_{c}=1$ pc and  metallicity $Z=Z_{\odot}$. Bottom panel: fixed values of $R_{c}=1$ pc, $M_{\textrm{1G}}=5 \times 10^{5}$ M$_{\odot}$ and different metallicities. The solid lines in all panels correspond to the case without feedback.}
    \label{fig2a}
\end{figure}

As shown by \cite{Silich2017,Silich2018,Silich2020, 2021MNRAS.504.3412L}, in the central zones of massive and compact star-forming clouds, the formation of a global wind is suppressed due to catastrophic gas cooling and the large intra-cluster gas pressures. However, at larger radii gas expulsion is possible given the lower ambient gas densities and pressures (see \citetalias{2021MNRAS.505.4669J} and \citealt{2021MNRAS.504.3412L}). The radius that separates these two zones (the super-wind radius $R_{\textrm{SW}}$), is determined by the relation (see \citetalias{2021MNRAS.505.4669J}):
\begin{equation}\label{WIND_RADIUS}
\frac{R_{\textrm{SW}}}{R_{c}}=\left\{
	\begin{array}{ll}
		1.70 \ln \left( 1.02 \beta^{-1} \right)^{0.58},  & \mbox{if }  \beta < 1,\\
		0 & \mbox{if } \beta \geq 1.
	\end{array}
\right.
\end{equation}
Here $\beta$ is given by:
\begin{equation}\label{WIND_RADIUS2}
\beta=\lambda \epsilon_{\textrm{1G}}^{1/3}\left(1-\epsilon_{\textrm{1G}} \right)^{1/3} \frac{R_{\textrm{c,pc}}^{5/3}}{M_{\textrm{gas},6}},
\end{equation}
where $M_{\textrm{gas},6}=M_{\textrm{gas}}/[10^{6} M_{\odot}]$, $R_{\textrm{c,pc}}=R_{\textrm{c}}/ [1$  $\textrm{pc}]$ and $\lambda=0.25, 0.40 \textrm{ and } 0.62$ for $Z/Z_{\odot}=10^{-2}, 10^{-1}$ and 1, respectively, where $Z_{\odot}$ is the solar metallicity.

The retained gas within the cluster $M_{\textrm{ret}}$ thus is $M_{\textrm{ret}}=M_{\textrm{gas}}\left(r \leq R_{\textrm{SW}}\right)$. Equations (\ref{WIND_RADIUS}-\ref{WIND_RADIUS2}) show that $M_{\textrm{ret}}$ is a function of the cluster parameters and of the star formation efficiency $\epsilon_{\textrm{1G}}$. Fig. \ref{fig2a} presents the retained gas to the initial cloud mass ratio $M_{\textrm{ret}}/M_{\textrm{tot}}$ as a function of $\epsilon_{\textrm{1G}}$. In all panels the solid lines show the case without feedback. The upper panel presents the $M_{\textrm{ret}}/M_{\textrm{tot}}$ ratio when the metallicity is solar and $M_{\textrm{1G}}=5 \times 10^{5}$ M$_{\odot}$. The dash, dotted and dash-dotted lines correspond to $R_{c}=1,2$ and 4 pc, respectively. Note that for the same star formation efficiency, the retained gas fraction is larger in clusters with smaller core radii due to stronger cooling caused by the larger ambient gas densities and pressures. For the same reason, increasing values of $M_{\textrm{1G}}$ lead to larger values of the retained gas, as shown in the middle panel of Fig. \ref{fig2a}. Finally, The bottom panel is the same but for a fixed core radius of $R_{c}=1$ pc and 1G mass $M_{\textrm{1G}}=5 \times 10^{5}$ M$_{\odot}$, while the metallicity changes as indicated by the legend. In this case, metal-poor clusters retain a larger fraction of their leftover gas because the stellar winds mechanical output decreases with the metallicity causing an ineffective gas expulsion despite the more limited influence of gas cooling in metal-poor clouds. 

Fig. \ref{fig2a} also shows that for a given set of cluster parameters ($M_{\textrm{1G}}, R_{c}, Z$), there exists a critical star formation efficiency $\epsilon_{\textrm{1G}}=\epsilon_{\textrm{winds}}$ above which $M_{\textrm{ret}}=0$. Here we are interested in high density star-forming clouds, where secondary stellar generations may be formed out of the gas left over from the 1G formation. Therefore, our 1G efficiencies never exceed the critical value $\epsilon_{\textrm{winds}}$. Larger 1G efficiencies $\epsilon_{\textrm{1G}}>\epsilon_{\textrm{winds}}$ would lead to complete gas expulsion and consequently no further star formation episodes would be possible.

The critical star formation efficiency $\epsilon_{\textrm{winds}}$ is determined by the condition $\beta=1$ in equation \ref{WIND_RADIUS2}, which leads to $R_{\textrm{SW}}=0$ and thus to complete gas expulsion ($M_{\textrm{ret}}=0$): 
\begin{equation}\label{ef1}
\epsilon_{\textrm{1G}}^{4/3} \left(1-\epsilon_{\textrm{1G}} \right)^{-2/3}-\frac{M_{\textrm{1G,6}}}{\lambda R_{\textrm{c,pc}}^{5/3}}=0.
\end{equation}
The fact that $M_{\textrm{gas,6}}=\left(\left(1- \epsilon_{\textrm{1G}}\right)/\epsilon_{\textrm{1G}} \right) M_{\textrm{1G,6}}$ has been used in equation \ref{WIND_RADIUS2} to derive \ref{ef1}. $\epsilon_{\textrm{winds}}$ is the numerical solution to equation \ref{ef1}, which provides an unique solution for each case of $M_{\textrm{1G}}$, $R_{c}$ and $Z$.

\subsection{Supernovae}
Following the literature, hereafter the cluster metallicity will be traced by the iron content $\textrm{[Fe/H]}$, for which we use the metallicity scale by \cite{2012MNRAS.426.1475U}.

The 1G core-collapse SN explosions occur during the first $3-40$ Myr of the cluster lifetime. Here, we consider high density clusters, with $\epsilon_{\textrm{1G}} < \epsilon_{\textrm{winds}}$ and $R_{\textrm{SW}}>0$. Then, a non-negligible fraction of the supernovae evolve within the leftover gas.

Supernova remnants (SNRs) that evolve within the retained gas ($r \leq R_{\textrm{SW}}$) face two possibilities. Either the supernovae are pressure confined in the central zone of the cluster polluting the leftover gas with iron-rich ejecta, or the supernovae experience blowout, in which the outermost section of the SNRs accelerate and fragment due to the density gradient and the metal-rich ejecta exits the cluster through the gaps between shell fragments (see \citetalias{2021MNRAS.505.4669J}). In such cases the SN products cannot enhance the leftover gas metallicity. 

The transition between these cases is given by the blowout radius $R_{\textrm{blow}}$ (see \citetalias{2021MNRAS.505.4669J}):
\begin{equation}
\frac{R_{\textrm{blow}}}{R_{c}}=\left\{
	\begin{array}{ll}
		1.4246 \left(\frac{M_{\textrm{gas,6}}}{R_{\textrm{c,pc}}} \right)^{0.35},  & \mbox{if    }  \frac{M_{\textrm{gas,6}}}{R_{\textrm{c,pc}}} \leq 2,\\
		1.5639\left(\frac{M_{\textrm{gas,6}}}{R_{\textrm{c,pc}}} \right)^{0.2},  & \mbox{if    }  \frac{M_{\textrm{gas,6}}}{R_{\textrm{c,pc}}} > 2.
	\end{array}
\right.
\label{blow_function}
\end{equation}

The leftover gas metallicity should remain within the observed limits. This puts another restriction on the possible star formation efficiency: $\epsilon_{\textrm{1G}} < \epsilon_{\textrm{iron}}$ since larger efficiencies would lead to $\Delta \textrm{[Fe/H]}>\Delta \textrm{[Fe/H]}_{\textrm{cri}}$, where $\Delta \textrm{[Fe/H]}_{\textrm{cri}}$ is the maximum observed iron spreads for GCs. 

$\epsilon_{\textrm{iron}}$ is determined by the leftover gas enrichment (e.g. \citetalias{2021MNRAS.505.4669J}; \citealt{2021MNRAS.506.4131W}):

\begin{equation}\label{metallicity_difference}
\Delta [\textrm{Fe/H}]  = \log \left(10^{[\textrm{Fe/H}]} + \frac {M_{\textrm{TSN}}} {Z^{\textrm{Fe}}_\odot  M_{\textrm{ret}}} \right)-[\textrm{Fe/H}],
\end{equation}
where $Z^{\textrm{Fe}}_\odot=0.0013$ \citep{2008MNRAS.391..354R} and $M_{\textrm{TSN}}$ is the iron mass added by the supernovae that explode within $R_{\textrm{blow}}$. $M_{\textrm{TSN}}$ is calculated upon the assumption that each SN provides $0.07$ M$_{\odot}$ of iron \citep[e.g.][]{2003ApJ...582..905H,2008MNRAS.391..354R}.

The critical efficiency $\epsilon_{\textrm{iron}}=\epsilon_{\textrm{iron}}\left(M_{\textrm{1G}}, R_{\textrm{c}}, [\textrm{Fe/H}] \right)$ is estimated numerically  with a precision of $10^{-4}$ in $\epsilon$ with equations \ref{WIND_RADIUS}, \ref{WIND_RADIUS2}, \ref{blow_function} and \ref{metallicity_difference}. 
\begin{figure}
\centering
	\includegraphics[width=1.0\columnwidth]{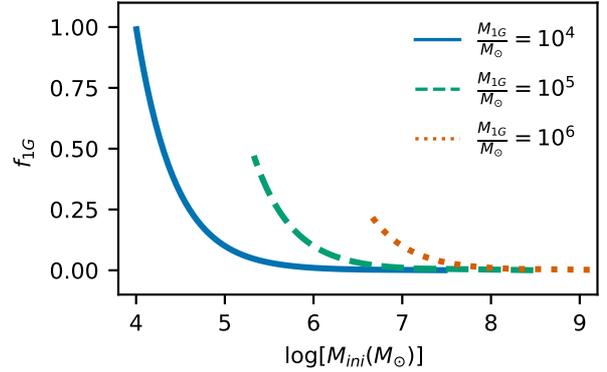}
        \caption{The fraction of 1G stars $f_{\textrm{1G}}$ as a function of the total initial cluster mass $M_{\textrm{ini}}$ for different values of the 1G mass $M_{\textrm{1G}}$, as shown in the legend. The curves are given by allowed models  with $\epsilon_{\textrm{1G}} \leq \epsilon_{\textrm{max}}$ (see equation \ref{eq_max}). Small values of $\epsilon_{\textrm{1G}}$ imply proportionally larger values of $M_{\textrm{ini}}$ and lower values of $f_{\textrm{1G}}$. }
    \label{fig6aux}
\end{figure}
\begin{figure*}
\centering
	\includegraphics[width=1.9\columnwidth]{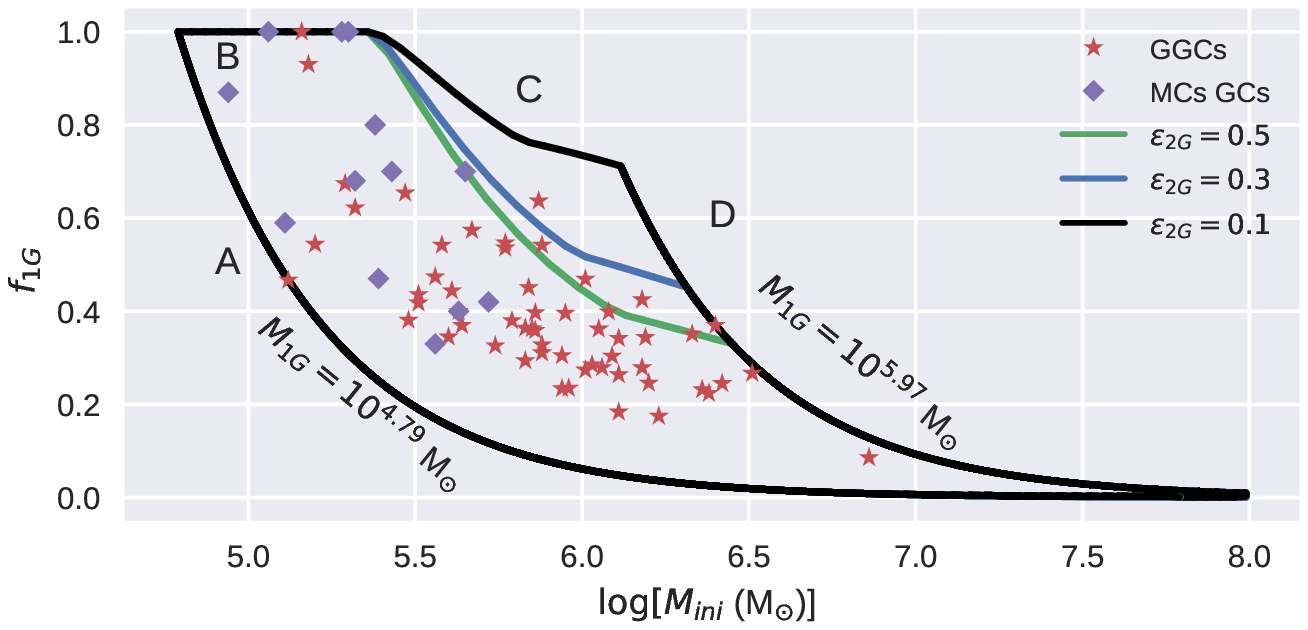}
        \caption{The fraction of 1G stars $f_{\textrm{1G}}$ as a function of the initial stellar cluster mass. The stars are observational results for GGCs, with $f_{\textrm{1G}}$ obtained by \citep{2017MNRAS.464.3636M} and initial cluster masses estimated by \citep{2019MNRAS.482.5138B}. The diamonds are results for the MCs globulars given by \citep{2019A&ARv..27....8G} with initial masses estimated as discussed in the text. Models results are contained within the areas bounded by the solid lines, which correspond to different values of $\epsilon_{\textrm{2G}}$, as indicated by the legend.}
    \label{fig6}
\end{figure*}
\section{The fraction of 1G stars in a GC as a function of the stellar mass}\label{eps_constraints}
As discussed above, stellar winds and supernovae feedback set two constraints on $\epsilon_{\textrm{1G}}$. First, $\epsilon_{\textrm{1G}}$ cannot exceed  $\epsilon_{\textrm{winds}}$ otherwise stellar winds would expel all the remaining gas from within the cluster. Second, $\epsilon_{\textrm{1G}}$ also has to be less than $\epsilon_{\textrm{iron}}$ to keep the iron spreads within the observational limits. Therefore, hereafter the upper limit for the 1G star formation efficiency $\epsilon_{\textrm{max}}$ is set as:
\begin{equation} \label{eq_max}
\epsilon_{\textrm{max}}=\textrm{min} \left(\epsilon_{\textrm{iron}}, \epsilon_{\textrm{winds}}\right).
\end{equation}
This ensures that models with $\epsilon_{\textrm{1G}} \leq \epsilon_{\textrm{max}}$ satisfy simultaneously both constraints, i.e.,  $f_{\textrm{1G}} \leq 1$ and $\Delta \textrm{[Fe/H]} \leq \Delta \textrm{[Fe/H]}_{\textrm{cri}}$. 

About $\sim 80 \%$ of the clusters in the sample of \cite{2017MNRAS.464.3636M} show negligible iron spreads  (see also \citealt{2009A&A...508..695C,2019ApJS..245....5B}). Therefore we select $\Delta \textrm{[Fe/H]}_{\textrm{cri}}=0.1$ dex in the following calculations. 

Equation \ref{eq_max} shows that the ability of the host molecular cloud to form MPs clusters with iron spreads that do not exceed the observed values, is determined by the set of input parameters: $M_{\textrm{1G}}$, $R_c$, $\textrm{[Fe/H]}$ and the 1G star formation efficiency $\epsilon_{\textrm{1G}}$. 
\DeclareRobustCommand*{\fdef}{\ensuremath{f_{\textrm{1G}}}}

The retained leftover gas enriched by trapped stellar ejecta can form one or several additional stellar populations, with a total mass $M_{\textrm{2G}}$:
\begin{equation}\label{eq_f1}
M_{\textrm{2G}}=\epsilon_{\textrm{2G}} M_{\textrm{ret}},
\end{equation}
where $\epsilon_{\textrm{2G}}$ is the second and next generations integrated star formation efficiency. 

The fraction of 1G stars thus is:
\begin{equation}\label{eq_f1b}
f_{\textrm{1G}}=\frac{M_{\textrm{1G}}}{M_{\textrm{ini}}},
\end{equation}
where:
\begin{equation}\label{eq_f2}
M_{\textrm{ini}}=M_{\textrm{1G}}+M_{\textrm{2G}},
\end{equation}
is the total initial stellar mass.

To obtain the possible values of the 1G fraction $f_{\textrm{1G}}$ and the total stellar mass $M_{\textrm{ini}}$ of the cluster, we first calculate $\epsilon_{\textrm{max}}=\epsilon_{\textrm{max}}\left( M_{\textrm{1G}}, R_{c}, \textrm{[Fe/H]}\right)$. Then, we run a number of models with star formation efficiencies $0 < \epsilon_{\textrm{1G}} \leq \epsilon_{\textrm{max}}$. The retained mass $M_{\textrm{ret}}=M_{\textrm{ret}}\left( R_{\textrm{c}}, M_{\textrm{1G}}, \textrm{[Fe/H]},\epsilon_{\textrm{1G}}\right)$ is calculated by means of equations \ref{WIND_RADIUS} and \ref{WIND_RADIUS2}. Finally, using equations \ref{eq_f1}-\ref{eq_f2}, one obtains a solution in the $f_{\textrm{1G}}$ vs $M_{\textrm{ini}}$ plane. 

Fig. \ref{fig6aux} presents 3 different cases for $M_{\textrm{1G}}/M_{\odot}=10^{4},10^{5}, 10^{6}$ with the remaining parameters fixed at $R_c=1.5$ pc, $\textrm{[Fe/H]}=-0.8$ dex and $\epsilon_{\textrm{2G}}=0.3$. Note that in the lowest mass case ($M_{\textrm{1G}}=10^{4}$ M$_{\odot}$, solid line in Fig. \ref{fig6aux}), the critical star formation efficiency is determined by $\epsilon_{\textrm{max}}=\epsilon_{\textrm{winds}}$. In this case, $f_{\textrm{1G}}=1$ when $\epsilon_{\textrm{1G}}=\epsilon_{\textrm{winds}}$, and then $f_{\textrm{1G}}$ drops from the left to the right with decreasing values of $\epsilon_{\textrm{1G}}$. Note that $f_{\textrm{1G}}\to 0 $ when $\epsilon_{\textrm{1G}} \to 0$, because in this case the 1G only accounts for a small fraction of the initial gas cloud and the stellar feedback is negligible ($M_{\textrm{1G}} \ll M_{\textrm{ret}} \sim M_{\textrm{tot}}$).

For larger values of $M_{\textrm{1G}}$, $\epsilon_{\textrm{winds}}$ becomes larger (see the middle panel of Fig. \ref{fig2a}). Therefore, in those cases it is the iron enrichment what sets the maximum allowed 1G star formation efficiency $\epsilon_{\textrm{max}}=\epsilon_{\textrm{iron}}<\epsilon_{\textrm{winds}}$. This explains why one always obtains $f_{\textrm{1G}}<1$ for the $M_{\textrm{1G}}=10^{5}$ and $10^{6}$ M$_{\odot}$ clusters. The maximum values of $f_{\textrm{1G}}$ in these cases are obtained for $\epsilon_{\textrm{1G}}= \epsilon_{\textrm{iron}}$ while lower values of $f_{\textrm{1G}}$ are obtained for smaller efficiencies. 
\begin{figure*}
\centering
	\includegraphics[width=1.9\columnwidth]{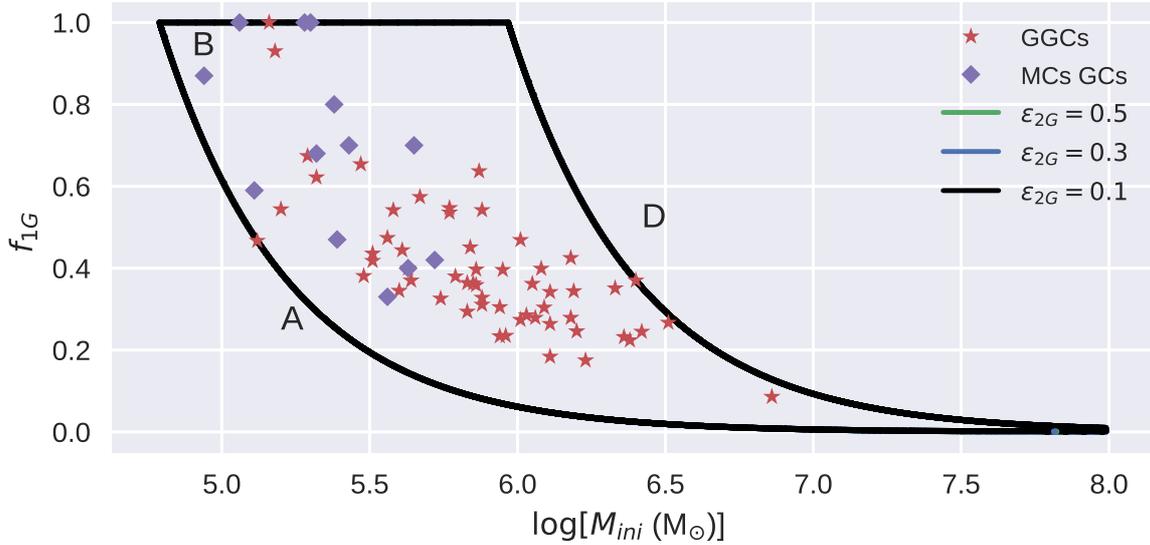}
        \caption{Same as Fig. \ref{fig6} but for core radii within the range $5$ pc $\leq R_{\textrm{c}} \leq$ $15$ pc.}
    \label{fig6b}
\end{figure*}
\section{Comparison of model-predicted $\lowercase{\textrm{f}}_{\textrm{1G}}$ values with observations}

Fig. \ref{fig6} presents the observed $f_{\textrm{1G}}$ vs $M_{\textrm{ini}}$ anti-correlation for GGCs (stars) and MCs globular clusters (diamonds). Initial $M_{\textrm{ini}}$ instead of present-day values of mass are used in the plot. The initial cluster masses for GGCs are the ones estimated by \cite{2019MNRAS.482.5138B} and for MCs clusters are approximated assuming that young (up to 2 Gyr old) and older GCs have lost the $10 \%$ and $30 \%$ of their initial mass, respectively \citep{2019A&ARv..27....8G}. The 1G fractions for the MCs globulars are compiled in Table 9 of \cite{2019A&ARv..27....8G} while the $f_{\textrm{1G}}$ data for GGCs are from \cite{2017MNRAS.464.3636M}. 

In order to compare the observed $f_{\textrm{1G}}$ vs $M_{\textrm{ini}}$ anti-correlation with our calculations, we run a set of models with input 1G masses $M_{\textrm{1G}}$ spanning an interval defined by the lower and upper limits of the data: 
\begin{equation}\label{M1G_range}
\textrm{min}\left\{f_{\textrm{1G}}M_{\textrm{ini}} \right\} \leq M_{\textrm{1G}} \leq \textrm{max}\left\{f_{\textrm{1G}}M_{\textrm{ini}} \right\},
\end{equation}
which are given by NGC 5466 ($\log \left[f_{\textrm{1G}}M_{\textrm{ini}}(M_{\odot}) \right] \sim 4.79$) and NGC 2419 ($\log \left[f_{\textrm{1G}}M_{\textrm{ini}}(M_{\odot}) \right] \sim 5.97$), respectively. We also assume the metallicities to be within $\textrm{[Fe/H]} \in [-1.5, -0.5]$ dex and core radii within $R_{c} \in [0.8,3]$ pc, which is expected if GCs were formed in dense environments \citep[e.g.][]{2014CQGra..31x4006K, 2015MNRAS.454.1658K}. Finally, $\epsilon_{\textrm{2G}}=0.1, 0.3$ and $0.5$ were selected in our calculations. 

Similar to Fig. \ref{fig6aux}, models with different $M_{\textrm{1G}}$ result in a set of curves in the $f_{\textrm{1G}}$ vs $M_{\textrm{ini}}$ plane, with all valid solutions enclosed by the solid line contours in Fig. \ref{fig6}. Here, different color contours correspond to different values of $\epsilon_{\textrm{2G}}$. Our calculations are in good agreement with the observational data as $f_{\textrm{1G}}$ can reach values as low as 0.1 and decreases with the star cluster mass.

The black contour ($\epsilon_{\textrm{2G}}=0.1$) in Fig. \ref{fig6} has been divided into several segments (A-D). The segments A and D are analogous to the solid and dotted lines presented in Fig. \ref{fig6aux}, but in this case the minimum and maximum $M_{\textrm{1G}}$ are given by equation \ref{M1G_range}. The segment B is composed by models formed with 1G efficiency $\epsilon_{\textrm{1G}}=\epsilon_{\textrm{winds}}$. For this reason $f_{\textrm{1G}}=1$ at this segment. Segment C results from models with $\epsilon_{\textrm{1G}}=\epsilon_{\textrm{iron}}$, which always have $f_{\textrm{1G}}<1$ as discussed above (see Fig. \ref{fig6aux}). All models enclosed by the solid line contours have 1G efficiencies below the maximum allowed value $\epsilon_{\textrm{1G}} <\epsilon_{\textrm{max}}$. Note that segments A, B and D are overlaid for models with different $\epsilon_{\textrm{2G}}$. The only difference among these cases is the location of segment C. 

Fig. \ref{fig6} shows that all but one observed cluster are located within the contour given by $\epsilon_{\textrm{2G}}=0.3$ and a considerable fraction are still in agreement with the area generated by the $\epsilon_{\textrm{2G}}=0.5$ calculations. Thus, our models require sufficiently large star formation efficiencies, as expected from those high pressure and density clouds \citep[e.g.][]{1997ApJ...480..235E}.

The 1G mass $M_{\textrm{1G}} \sim 2 \times 10^{5}$ M$_{\odot}$ (the crossing point between segments B and C in Fig. \ref{fig6}) splits models with $\epsilon_{\textrm{max}}=\epsilon_{\textrm{winds}}$ from those with $\epsilon_{\textrm{max}}=\epsilon_{\textrm{iron}}$. Hence, $M_{\textrm{ini}} \sim 2 \times 10^{5}$ M$_{\odot}$ is a critical mass for GGCs and MCs globulars as more massive clusters always host MPs. 

The data constraints the values of $M_{\textrm{1G}}$ and $\textrm{[Fe/H]}$ that were used as input values in the calculations presented in Fig. \ref{fig6}. However, for the core radius we have assumed that GGCs and MCs GCs were compact at their formation ($0.8$ pc $\leq R_{\textrm{c}} \leq$ $3$ pc). Fig. \ref{fig6b} presents the results for more extended clusters ($5$ pc $\leq R_{\textrm{c}} \leq$ $15$ pc). As densities and pressures drop with the cluster core radius (see equations \ref{eq1} and \ref{eq2}), stellar winds feedback is stronger (see also the top panel in Fig. \ref{fig2a}). Hence, unlike Fig. \ref{fig6} (where $\epsilon_{\textrm{max}}=\epsilon_{\textrm{iron}}$ for $M_{\textrm{1G}} \gtrsim 2 \times 10^{5}$ M$_{\odot}$), the uppermost section (segment B with $\epsilon_{\textrm{1G}}=\epsilon_{\textrm{winds}}$ and $f_{\textrm{1G}}=1$) of the black contour in Fig. \ref{fig6b} extends to all the $M_{\textrm{1G}}$ range. Note that the contours presented in Fig. \ref{fig6b} will be the same for larger core radii, as the above arguments hold for even more extended clouds. 

The observed $f_{\textrm{1G}}$ vs $M_{\textrm{ini}}$ anti-correlation is in	better agreement with the compact proto-globular cluster models presented in Fig. \ref{fig6} compared with those shown in Fig. \ref{fig6b}. Not only because they follow closely the observed data points (specially for the $\epsilon_{\textrm{2G}}=0.5$ or 0.3 cases), but also because they, in agreement with observations, do not predict the existence of massive ($M_{\textrm{ini}} \sim 10 ^{6}$ M$_{\odot}$) SSPs clusters. Therefore, our results favor a scenario where young GCs formed as compact and dense structures. This is predicted by recent theoretical models \citep{2015MNRAS.454.1658K, 2021MNRAS.507.5492C} and it is also reported for YMCs in nearby galaxies \citep[e.g.][]{2004A&A...416..537L, 2020MNRAS.492..993C, 2021MNRAS.508.5935B}. Furthermore, cluster survival is more likely for compact and massive clusters, hence suggesting the old GCs should have been compact at their birth to endure a Hubble time of dynamical evolution \citep[e.g.][]{2021MNRAS.500.4422C}. 

\begin{figure}
\centering
	\includegraphics[width=1.0\columnwidth]{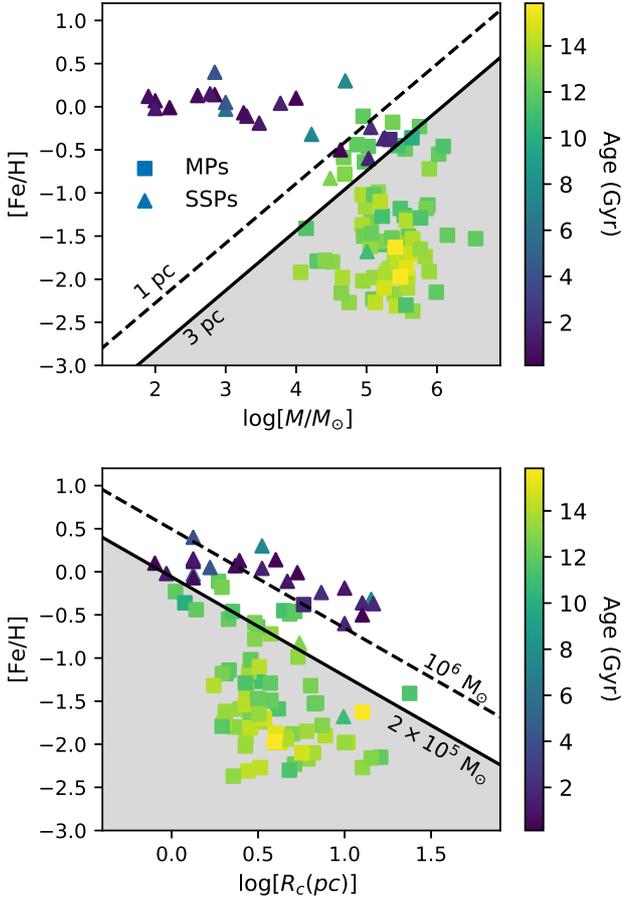}
        \caption{The metallicity $\textrm{[Fe/H]}$ as a function of the cluster mass (upper panel) and cluster core radius (bottom panel) for Galactic and extragalactic clusters. Squares and triangles indicate that MPs are clearly present/absent within the cluster, respectively. The color of each data point presents its estimated age (indicated by the color bar). See the text for a discussion on the data. The dashed and solid lines show the critical metallicity $\textrm{[Fe/H]}_{\textrm{crit}}$ (see equation \ref{eq_metal_core_mass}) for core radii $R_{c}=1$ and 3 pc (top panel) and cluster masses $10^{6}$ M$_{\odot}$ and $10^{5}$ M$_{\odot}$ (bottom panel), respectively. The shaded areas correspond to cases where only MPs can exist for $R_{c}=3$ pc (top panel) and $2 \times 10^{5}$ M$_{\odot}$ (bottom panel).}
    \label{fig3}
\end{figure}

\subsection{Global parameters and the onset of $\textrm{MP}\lowercase{\textrm{s}}$}

MCs globulars provided new insights to the MPs problem. Indeed, MPs were discovered in NGC 1978 \citep{2018MNRAS.477.4696M}, and Hodge 6 \citep{2019MNRAS.484.4718H}, hosting some of the smaller fractions of enriched stars ($f_{\textrm{1G}} \gtrsim 0.8$) within the Galactic and extra-galactic clusters studied up to date. However, NGC 419 \citep[e.g.][]{2017MNRAS.468.3150M, 2020ApJ...893...17L}, NGC 1806 \citep{2014ApJ...793L...6M} and NGC 1783 \citep{2018ApJ...853..186Z} were found to be consistent with SSPs clusters. The fact that these clusters likely had similar initial masses and that MPs have not been found in young massive clusters (YMCs) \citep{2016MNRAS.460.1869C}, led several authors to argue that cluster age may also be an indicator of the presence of MPs within clusters and that only clusters older than $\sim 2$ Gyr show evidences of MPs \citep{2019ApJ...876...94L, 2019MNRAS.487.5324M, 2021MNRAS.505.5389M}. However, recently \cite{2021arXiv211206964C} showed, by means of UV HST and optical photometry of main sequence stars (MS), instead of previous studies based on red giant branch stars (RGB), that the $\sim 1.5$ Gyr old cluster NGC 1783 does host MPs. Hence, although MPs may be common among clusters regardless of their age, given their similar masses the MCs clusters still indicate that an additional parameter(s) besides mass, determine the strength of the MPs phenomenon within GCs. Our results show that for the mass range of MCs GCs ($\sim 10^{5}$ M$_{\odot}$), a large intrinsic scatter (see Fig. \ref{fig6}) is expected due to variations in metallicity, cluster radius and star formation efficiency, the other parameters besides mass which determines the presence of MPs.

\section{Impact of metallicity on the formation of $\textrm{MP}\lowercase{\textrm{s}}$ clusters}
The stellar winds mechanical output, gas cooling and the chemical enrichment of the leftover gas are all metallicity-dependent processes. Indeed, smaller iron spreads are obtained for metal-rich star-forming clouds (see equation \ref{metallicity_difference}). Equations (\ref{WIND_RADIUS}-\ref{WIND_RADIUS2}) also show that, for a given 1G mass $M_{\textrm{1G}}$ and core radius $R_c$, large metallicities promote gas expulsion (see the bottom panel in Fig. \ref{fig2a}). Therefore, above some critical metallicity, $\textrm{[Fe/H]}_{\textrm{crit}}$, the maximum allowed 1G star formation efficiency is determined by the winds mechanical feedback, $\epsilon_{\textrm{max}}=\epsilon_{\textrm{winds}}$, while in cases with metallicities smaller than this critical value, $\epsilon_{\textrm{max}}$ is given by the iron enrichment constraint, $\epsilon_{\textrm{max}}=\epsilon_{\textrm{iron}}$. Hence, MPs must be present in all clusters with metallicities $\textrm{[Fe/H]} \leq \textrm{[Fe/H]}_{\textrm{crit}}$.

We estimated numerically $\textrm{[Fe/H]}_{\textrm{crit}}$ as a function of the cluster core radius and mass and found that it is well fitted by:
\begin{equation}\label{eq_metal_core_mass}
\textrm{[Fe/H]}_{\textrm{crit}}=\log \left[ \left(\frac{M_{\textrm{1G}}} { 10^{4}M_{\odot}}\right)^{0.69} \left( \frac{R_{c}}{\textrm{pc}} \right)^{-1.15}\right]-0.89 \textrm{ dex}.
\end{equation}
The critical metallicity is shown in Fig. \ref{fig3} for core radii $R_{c}=1$ (dashed line) and 3 pc (solid line) as a function of mass in the top panel, and for cluster masses $10^{6}$ M$_{\odot}$ (dashed line) and $2 \times 10^{5}$ M$_{\odot}$ (solid line) as a function of the core radius in the bottom panel. Squares and triangles display Galactic and extragalactic clusters with MPs and SSPs, respectively. The color of each data point presents the estimated cluster age (see the color bar). For GGCs, masses and radii are taken from \cite{2018MNRAS.478.1520B}, while the cluster ages and metallicities are taken from the compilation of \cite{2017A&A...607A..44B}. For the MCs clusters, data on mass, age and metallicity are from \cite{2017A&A...607A..44B}. The mass and age of NGC 1978 and mass, metallicity and age of NGC 1783 are from \cite{2019A&ARv..27....8G}. Cluster radii for the MCs clusters and data for OCs and the Fornax dwarf clusters are taken from \cite{2016A&A...587A..53K}. Finally, whether or not the GGCs and MC globulars host MPs is from \cite{2017A&A...607A..44B}.

Fig. \ref{fig3} must be taken with care as present-day instead of initial cluster masses and radii are used in this figure. However, even though the cluster mass and radius evolve due to several dynamical processes \citep[e.g.][]{2011MNRAS.413.2509G, 2018MNRAS.481..268T}, it is likely that young clusters (due to their age) and very massive clusters (due to their mass) have structural parameters similar to their initial values \citep[e.g.][]{2015MNRAS.454.1658K}.

Note that each panel in Fig. \ref{fig3} is split into two regions. SSPs clusters should all be located above the critical lines. Hence, the solid lines are in better agreement with the available data as all (with the exception of Rup 106) clusters hosting SSPs are located above the solid lines in Fig. \ref{fig3} and $\sim 90 \% $ of the clusters hosting MPs are located bellow the solid lines in Fig. \ref{fig3}. The shaded zones correspond to cases where, according to our calculations, only MPs can exist. It is likely that the model predictions will be in even better  agreement with the observations if one makes use of the initial rather than the current star cluster parameters as in this case the oldest clusters (see the color scale in the legend), should be located further to the right in the top panel and further to the left in the bottom panel. Indeed, if only young and intermediate age clusters ($<8$ Gyr) were considered the agreement reaches $100 \%$. 


\section{SUMMARY AND CONCLUSIONS}\label{summarySection}
Since the discovery of multiple stellar populations in globular clusters \citep[e.g.][]{ 1999Natur.402...55L, 2004ApJ...605L.125B, 2009A&A...505..139C, 2012A&A...544A..12G, 2015A&A...578A.116C, 2015AJ....149...91P,2015MNRAS.447..927M, 2015ApJ...808...51M, 2019MNRAS.487.3815M} numerous ideas have been proposed to explain the sequence followed to end up with multiple stellar generations  despite stellar feedback, or the possibility of star formation in places already occupied by a massive burst of star formation (see section \ref{introductionI}).

We have presented here another idea which results from considering star formation at very high densities and have compared our results with observational results. In our calculations, for every given proto-globular cluster cloud with mass $M_{\textrm{tot}}$ and metallicity $\textrm{[Fe/H]}$, the star formation efficiency of the first stellar generation ($\epsilon_{1G}$ < 1) defines the 1G mass and the amount of primordial gas left over from the star formation event, with  both components sharing the metallicity of the primordial cloud. For both of these components we assumed a Gaussian distribution with a core radius ($R_c$). 

By means of a standard IMF, we found for each model the local impact from the winds from massive stars and work out for which of these models, if not all, winds are either capable of totally or partially, disrupting the gas left over from star formation or become trapped within the high pressure central regions. The value of $\epsilon_{1G}$ also defines the number of expected supernovae. Many of them (occurring in the outskirts of the cluster) should take place within the cluster wind and thus contribute to its power as their ejecta streams away from the cluster. Other explosions, closer to the center, will either experience blowout and in this way expel their metals out of the cluster or be confined by the surrounding pressure, mixing their ejecta with the surrounding gas while enhancing its metallicity.

Metallicity is one of the major issues in our calculations which consider in full detail the enhancement acquired by the gas leftover from the formation of a 1G, leaving it ready to experience a second burst, a 2G with the correct metallicity. For a given set of structural parameters ($R_{c}, M_{\textrm{1G}}$ and $\textrm{[Fe/H]}$), we have shown that a key parameter is the 1G star formation efficiency, as it regulates both the ability of the proto-cluster to retain leftover gas and the iron enrichment by SN ejecta. 

The 1G fraction ($f_{\textrm{1G}}$) becomes equal to 1 for cases where stellar winds clear all the leftover gas, those with $R_{\textrm{SW}}=0$, while $f_{\textrm{1G}}$ acquires smaller values for multiple population clusters where winds are ineffective either due to a low value of $\epsilon_{\textrm{1G}}$ or because of the strong cooling that occurs in high density clusters, where iron enrichment sets the 1G star formation efficiency. Note that $f_{\textrm{1G}}$ can reach the value of  0.1 as in the observational data. Fig. \ref{fig6} shows that all observed clusters lie within the boundaries established by the theoretical constraints and globally the resultant $f_{\textrm{1G}}$ is inversely proportional to the log of the total cluster mass, as demanded by the observations.

The major implications from the good agreement between our calculations and the observations are that proto-globular cluster clouds should initially have presented structural parameters which make them massive and compact and also able to withstand the mechanical impact from the strong winds and SN from a first stellar generation. Such clouds should have been able to trap the sufficient supernovae products to enhance their metallicity and thus of all secondary stellar generations. 

Our calculations were also compared with recent observational results (see Fig. \ref{fig3}) which  indicate the run of metallicity and age as a function of the cluster mass or cluster radius, for Galactic and extragalactic globular clusters. Such diagrams indicate the conditions required for MPs clusters. It is clear that the mass of the clusters is an important parameter leading to multiple population clusters. However, other parameters, such as the dimension or core radius, the metallicity and the 1G star formation efficiency are also important parameters in the formation of MPs clusters.

\section*{Acknowledgements}
We thank our anonymous referee for the useful suggestions that improved the presentation of our results. This study was supported by CONACYT-M\'exico research grant A1-S-28458. SJ acknowledge the support by CONACYT-M\'exico (scholarship registration number 613136). The authors thankfully acknowledge the support provided by the Laboratorio Nacional de Superc\'omputo del Sureste de M\'exico, CONACYT member of the network of national laboratories.




\section*{DATA AVAILABILITY STATEMENT}
No new data were generated in support of this research. Datasets analyzed during the current study are openly available in the cited literature.

\bibliographystyle{mnras}

\bibliography{ref}


\appendix
\section{The impact of mass segregation}\label{Ap1}
\begin{figure}
\centering
	\includegraphics[width=1.0\columnwidth]{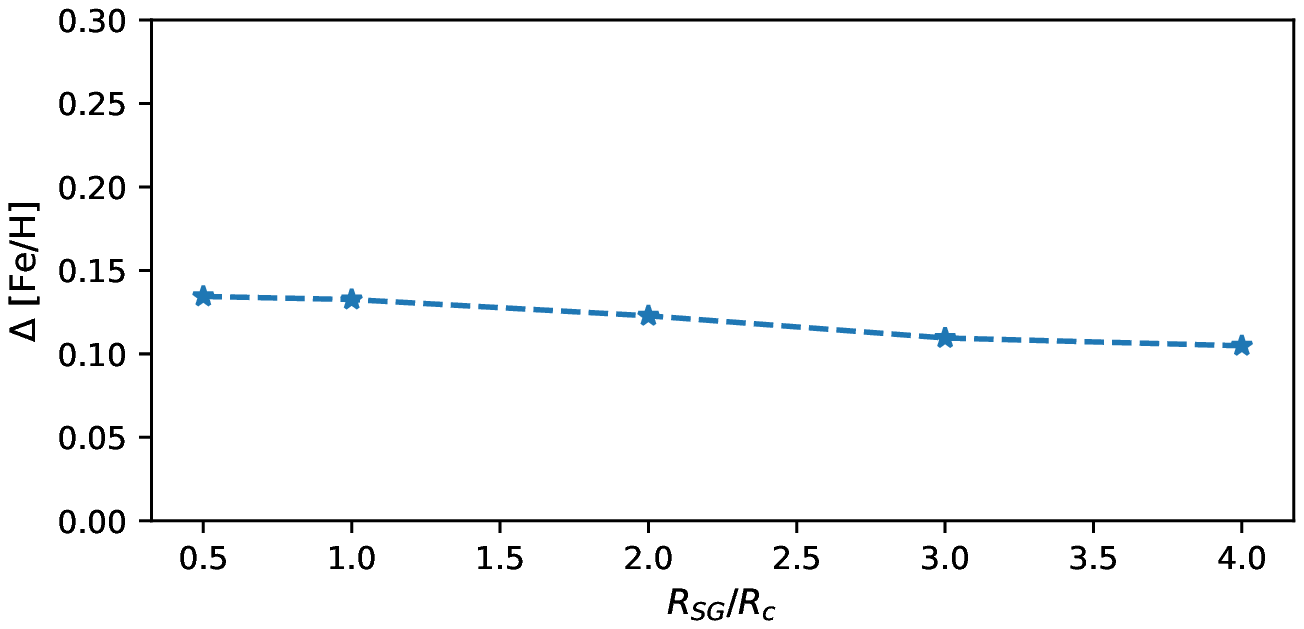}
        \caption{The metallicity enhancement $\Delta \textrm{[Fe/H]}$ as a function of $R_{\textrm{SG}}/R_{c}$ for a model with $M_{\textrm{1G}}=10^{5}$ M$_{\odot}$, $\epsilon_{\textrm{1G}}=0.05$, $R_{c}=1.5$ pc and $\textrm{[Fe/H]}=-1$ dex. }
    \label{fig_AP1}
\end{figure}

The superwind, $R_{\textrm{SW}}$, and blowout, $R_{\textrm{blow}}$, radii that determine the leftover gas metallicity and the amount of mass available for a second and next episodes of star formation fall in the range $(1.5-3)R_{c}$ for most of our models (see Table 2 in \citetalias{2021MNRAS.505.4669J}). Most of the stellar mass (up to $90\%$) is located within this range even if the massive stars are located throughout the cluster volume as assumed in our calculations. Indeed, the number of massive stars as a function of radius is:

\begin{equation}\label{eqA1}
N_{\textrm{massive}}\left(r \right)=\left\{
	\begin{array}{ll}
		\alpha \exp [-\frac{1}{2}r^{2}/R_{c}^{2}],  & \mbox{if }  r \leq  R_{\textrm{SG}},\\
     0 & \mbox{if } r > R_{\textrm{SG}},
	\end{array}
\right.
\end{equation}
with:
\begin{equation}
\alpha=\frac{N_{\textrm{tot}}}{\left( 2 \pi \right)^{3/2} R_{c}^{3}\left[ \textrm{erf} \left( \frac{R_{\textrm{SG}}}{\sqrt{2} R_{c}}\right) -\sqrt{\frac{2}{\pi}}\frac{R_{\textrm{SG}}}{R_{c}} \exp \left( -\frac{R_{\textrm{SG}}^{2}}{2 R_{c}^{2}}\right)\right] },
\end{equation}
where $N_{\textrm{tot}}$ is the total number of massive stars and $R_{\textrm{SG}}$ is the mass segregation radius. One can obtain $\Delta\textrm{[Fe/H]}$ for different mass segregation radii by making use these equations, equation \ref{blow_function} for $R_{\textrm{blow}}$ and calculating $R_{\textrm{SW}}$ numerically (see \citetalias{2021MNRAS.505.4669J}). Fig. \ref{fig_AP1} presents the results of such calculations in the case when $M_{\textrm{1G}} = 10^5$ M$_{\odot}$, $\epsilon_{\textrm{1G}} = 0.05$, $R_c = 1.5$ pc and $\textrm{[Fe/H]} = -1$ dex. As it is expected, $R_{\textrm{SW}}$ decreases and a larger number of SNe can be retained in models with smaller $R_{\textrm{SG}}$. However, there is only a small ($\sim 0.03$ dex) difference between models with $R_{\textrm{SG}} = 4 R_c$ and $R_{\textrm{SG}} = 0.5 R_c$. Thus the impact of mass segregation on our results is negligible.

\bsp	
\label{lastpage}
\end{document}